\documentclass[nofootinbib,aps,a4paper,letterpaper,superscriptaddress,
twocolumn,times,showpacs,eqsecnum]{revtex4}

\usepackage{amsmath}
\usepackage{amsfonts}
\usepackage{booktabs}
\usepackage{multirow}
\usepackage{siunitx}
\usepackage{adjustbox}
\pdfoutput=1
\usepackage{graphicx}
\usepackage{color}
\usepackage{braket}
\usepackage{dcolumn}
\usepackage{bm,url}
\usepackage[linktocpage]{hyperref}
\usepackage{subfigure}
\usepackage{amsfonts}
\usepackage[usenames,dvipsnames,svgnames]{xcolor}
\usepackage{hyperref}
\definecolor{oxfordblue}{rgb}{0.0, 0.13, 0.28}
\definecolor{burgundy}{rgb}{0.5, 0.0, 0.13}
\definecolor{darkolivegreen}{rgb}{0.33, 0.42, 0.18}
\definecolor{darkblue}{rgb}{0,0,0.5}
\definecolor{richcarmine}{rgb}{0.84, 0.0, 0.25}
\definecolor{darkblue}{rgb}{0,0,0.5}
\definecolor{bluer}{rgb}{0.00,0.50,0.75}{}
\hypersetup{colorlinks=true, citecolor=red, linkcolor=blue,
 urlcolor = magenta, filecolor=magenta}

\newcommand{\mpl}{M_{\rm P}}
\newcommand{\mplfour}{M_4}
\newcommand{\mplfive}{M_5}
\newcommand{\lam}{\lambda}

\begin{document}

\title{Swampland diagnostics and CMB bounds on high-energy brane chaotic
inflation}

\author{Saeed Noori Gashti
}
\email{sn.gashti@du.ac.ir}
\affiliation{School of Physics, Damghan University, P. O. Box 3671641167, Damghan,
Iran}
\affiliation{Center for Theoretical Physics, Khazar University, 41 Mehseti Street,
Baku, AZ1096, Azerbaijan}

\author{Behnam Pourhassan}
\email{b.pourhassan@du.ac.ir}
\affiliation{School of Physics, Damghan University, P. O. Box 3671641167, Damghan,
Iran}
\affiliation{Center for Theoretical Physics, Khazar University, 41 Mehseti Street,
Baku, AZ1096, Azerbaijan}

\author{\.{I}zzet Sakall{\i}}
\email{izzet.sakalli@emu.edu.tr}
\affiliation{Physics Department, Eastern Mediterranean University, Famagusta 99628,
Cyprus}

 \author{Emmanuel N. Saridakis}
 \email{msaridak@noa.gr}
 \affiliation{Institute for Astronomy, Astrophysics, Space Applications and
Remote Sensing, National Observatory of Athens, 15236 Penteli, Greece}
 \affiliation{Departamento de Matem\'{a}ticas, Universidad Cat\'{o}lica del
  Norte, Avda. Angamos 0610, Casilla 1280, Antofagasta, Chile}
 \affiliation{CAS Key Laboratory for Research in Galaxies and Cosmology,
School
  of Astronomy and Space Science,
  University of Science and Technology of China, Hefei 230026, China}

\begin{abstract}
High-energy Randall-Sundrum type-II inflation modifies the relation between
the inflaton potential, expansion rate, and tensor amplitude,
changing the tension between simple inflation and Swampland criteria. We
reassess monomial brane inflation, we derive its leading-order scalar and tensor
predictions, and we compare them with Planck and ACT-era constraints. We find that
the scalar
tilt, tensor amplitude, and e-fold number obey a simple relation. Moreover, we
show that the four-dimensional relations between CMB
observables and potential derivatives fail in the high-energy regime, where
they also involve the energy scale relative to the brane tension. Thus,  the
tensor amplitude alone does not determine the potential slope.
The scalar weak gravity conjecture and its strong version yield exact,
finite field-space intervals, which the scalar normalization translates
into brane-tension bounds. Higher-tilt ACT-era data allow a broader range
of monomial powers below unity than Planck, whose compatible region is
concentrated at smaller e-fold numbers. Additionally, sub-Planckian excursion and the
local Swampland criteria can hold together in a finite region that meets the
high-energy condition. Finally, the trans-Planckian censorship conjecture is far
more restrictive, namely  alongside  the lower limit on the brane tension from
five-dimensional gravity  it
leaves only a corner with about 32 e-folds and powers below 0.04. Our results
identify phenomenological compatibility within the effective
brane description without establishing a complete ultraviolet realization.
\end{abstract}


\maketitle

\section{Introduction}

Inflationary model building connects precision cosmology with issues related to
the ultraviolet consistency of effective field theories. Current observations
constrain the scalar spectral index with increasing precision, while the
tensor-to-scalar ratio remains bounded from above
\cite{Planck2018Inflation,BK18}. At the same time, the
Swampland framework has proposed conjectural criteria that may distinguish
effective theories compatible with quantum gravity from those that are not
\cite{k,o,bbb}. Nevertheless, their application to inflation is particularly
challenging.
The distance conjecture constrains large field excursions, while the original
de Sitter conjecture requires a sufficiently steep potential. These
requirements can be hard to reconcile with the predictions of simple
single-field inflation.

The Randall-Sundrum type-II (RS-II) braneworld \cite{Randall:1999vf} provides a
framework in which
the inflationary dynamics are modified by the embedding of our four-dimensional
Universe in a five-dimensional bulk \cite{Binetruy:1999ut,Maartens:1999hf}. In such
a model the effective Friedmann equation contains
a term quadratic in the energy density, and hence, contrary to
standard four-dimensional cosmology, at high energies  the Hubble
rate is   proportional to the potential rather than to its square root, a feature
that leads to interesting behavior  \cite{Kim:1999ja, Copeland:2000hn, Piao:2001dd,
Bento:2001hu, Bento:2002np, Liddle:2003zw, Liddle:2003gw, Seery:2003ge,
Calcagni:2003sg, Tsujikawa:2003zd, Buchel:2004qg, Tsujikawa:2004dm, Calcagni:2004bb,
Copeland:2005qe, Campuzano:2005us, Lidsey:2006md, Saridakis:2007cf, Saridakis:2007ns,
Herrera:2010vv, Wong:2010rg, Nozari:2012cy, Calcagni:2013lya, Kofinas:2014qxa,
Choudhury:2014sua, Neupane:2014vwa, Herrera:2015aja, Barros:2015evi, Bilic:2018uqx,
Adhikari:2020xcg}. This changes the relations among the
potential slope, the slow-roll parameters, the field excursion, and the tensor
amplitude. These features motivated the original study of chaotic inflation
with a monomial potential, $V(\phi)=A\phi^p$, on the brane
\cite{LinNgCheung2019}, an  analysis that showed how brane corrections could
permit a sub-Planckian field excursion while allowing a steeper potential.

However, the observational setting has since changed. In particular, Planck 2018
reported
$n_s=0.9649\pm0.0042$ and, in combination with BICEP/Keck BK15, the bound
$r_{0.002}<0.056$ \cite{Planck2018Inflation,BK18}. ACT DR6 gives
$n_s=0.9666\pm0.0077$ for ACT alone, while combinations with larger-scale
Planck data, CMB lensing, and DESI Year 1 BAO favor a higher tilt,
$n_s\simeq0.974\pm0.003$ \cite{ACTDR6,ACTExtended}. One can also consider the
combined ACT, Planck, and BICEP/Keck benchmark, namely
$n_s=0.9743\pm0.0034$ and $r_{0.05}<0.038$
\cite{ACTExtended,ACTInflationBenchmark2025,ACTInflationBenchmark2026}. These
likelihood combinations are distinct, and in the following  we use them separately.

On the other hand, the Swampland literature has expanded beyond the original
distance and
de Sitter conjectures. Later proposals include the refined and
further-refined de Sitter criteria, as well as the scalar weak gravity
conjecture and its strong version
\cite{AndriotRoupec2019,GonzaloIbanez2019}. These criteria have been
applied to inflation and dark energy in a range of settings, including
warm inflation, eternal inflation, and axion models
\cite{50,51,52,53,54,55,56,57,58,59,60,61,62}. They are considered as
complementary diagnostics, since they address distinct aspects of
effective-field-theory consistency. Moreover, the trans-Planckian
censorship conjecture provides a separate constraint on the inflationary
scale and, in a brane setting, on the brane tension
\cite{BedroyaVafa2019,MohammadiGolanbariNasriSaaidi2022,Li:2025cxn}.

Having these in mind, in this work we reassess monomial inflation in the
high-energy
RS-II regime. In particular, we derive its leading-order scalar and tensor
predictions
and show why the five-dimensional tensor normalization changes the
relations used to infer potential derivatives from CMB observables.
Consequently, the tensor amplitude alone does not determine the potential
slope. We evaluate various criteria, such as the de Sitter, refined and
further-refined de Sitter ones, as well as
the scalar weak gravity
conjecture (SWGC), the strong scalar weak gravity
conjecture (SSWGC), the distance conjecture, and the trans-Planckian
censorship conjecture (TCC), and we combine them with
CMB constraints and the scalar-amplitude normalization. This procedure
yields bounds on the monomial power, the e-fold number, and the brane
tension. Additionally, we account for reheating, since it affects the
e-fold number, particularly for fractional powers.

Our conclusions apply to the brane effective description. The inflaton is
a canonical scalar localized on the brane, therefore its potential derivatives
are well defined. However, evaluating four-dimensional field-space
criteria in a five-dimensional background does not establish a complete
ultraviolet construction. Such a construction would require information
about the bulk, stabilization, and the relevant tower of states. Furthermore, tensor
perturbations carry a bulk normalization, while
scalar fluctuations can couple to bulk metric modes at high energies
\cite{KoyamaEtAl2007}. Nevertheless, these qualifications do not prevent
us from assessing the phenomenological viability of the effective model, they just
  define the goal of the analysis  and distinguish it from a claim
of ultraviolet completion.

Recent efforts to address the higher scalar tilt indicated by particular CMB
data have also considered anomalous inflaton scaling, scale-independent
or Weyl-invariant $R^2$ sectors, heavy-field effects, non-canonical scalar fields,
and other curvature
or radiative corrections
\cite{Kouniatalis:2025orn, Stavros:2026evy, Kumar:2025apf, Karydas:2021wmx,
Lola:2020lvk,R2ScaleIndependent,WeylR2EmergentPlanck,WeylScalingR2,ACTWeyl2025,
HeavyFieldACT2025,PolynomialACT2025,RadiativeHiggsACT2025,OikonomouR2ACT2025,
OikonomouStringACT2025,KetovFRACT2025,OdintsovEGB2025,AhmedRehmanACT2025,
AddaziKetovACT2025,ZharovSobolACT2025,YuennanACT2026,ZhangZhaiChenACT2025,
MohammadiWangACT2025,YiACT2025,GashtiHiggsPalatini2025,AfsharACT2026,
Saridakis:2026vxo}.
These studies provide broader observational context, while the present
work focuses on monomial inflation in the RS-II braneworld.

The paper is organized as follows. Section \ref{Highenergybrane} presents the
high-energy
brane background and the monomial slow-roll dynamics, including the scalar
and tensor observables. Section \ref{Swamplanddiagnostics} develops the Swampland
diagnostics for
the monomial potential. Section \ref{CMBconstraints} confronts the model with CMB
benchmarks
and examines the joint observational and Swampland viability. Section
\ref{Theoreticalconsistency}
discusses the validity of the approximations, the relation to previous
brane-Swampland studies, and the limitations of the analysis. Finally, Section
\ref{Conclusions}
presents our conclusions.

\section{High-energy brane inflationary dynamics}
\label{Highenergybrane}

In this section we first establish the RS-II brane background and specify the
high-energy regime in which the quadratic energy-density correction dominates the
expansion. Then we derive the slow-roll predictions for a monomial potential and
discuss the tensor relations needed for the subsequent Swampland and CMB analyses.

\subsection{Brane background and high-energy regime}

In the Randall-Sundrum type-II (RS-II) scenario, our four-dimensional Universe is
described by a three-brane embedded in a five-dimensional anti-de Sitter bulk. For a
homogeneous and isotropic brane, the effective Friedmann equation takes the form
\cite{Randall:1999vf,Binetruy:1999ut,Maartens:1999hf}
\begin{equation}
H^2=\frac{\Lambda_4}{3}
+\frac{\rho}{3\mpl^2}\left(1+\frac{\rho}{2\lam}\right)
+\frac{\mathcal C}{a^4},
\label{eq:fullFriedmann}
\end{equation}
with $H$ the four-dimensional Hubble function, and where $\Lambda_4$ is the effective
four-dimensional cosmological constant, $\lam$ is
the brane tension, and the term $\mathcal C/a^4$ describes the dark-radiation
contribution associated with the bulk Weyl sector. During inflation, the effective
cosmological constant and dark radiation are neglected. The Friedmann equation then
reduces to \cite{LinNgCheung2019,MohammadiGolanbariNasriSaaidi2022}
\begin{equation}
H^2=\frac{\rho}{3\mpl^2}\left(1+\frac{\rho}{2\lam}\right).
\label{eq:braneF}
\end{equation}

For a canonical scalar field $\phi$ confined to the brane, the energy density and
pressure are
\begin{equation}
\rho_\phi=\frac{1}{2}\dot{\phi}^{\,2}+V(\phi),
\qquad
P_\phi=\frac{1}{2}\dot{\phi}^{\,2}-V(\phi),
\end{equation}
and the field equation is
\begin{equation}
\ddot{\phi}+3H\dot{\phi}+V_{,\phi}=0.
\label{eq:KG}
\end{equation}
In the slow-roll regime, $\dot{\phi}^{\,2}\ll V$ and
$|\ddot{\phi}|\ll 3H|\dot{\phi}|$, and thus equations
\eqref{eq:braneF} and \eqref{eq:KG}   become
\begin{equation}
H^2\simeq\frac{V}{3\mpl^2}
\left(1+\frac{V}{2\lam}\right),
\qquad
3H\dot{\phi}\simeq -V_{,\phi}.
\label{eq:SRbrane}
\end{equation}

We proceed by  defining  the useful  dimensionless quantity
\begin{equation}
u\equiv\frac{V}{2\lam},
\label{eq:uDef}
\end{equation}
and hence the high-energy regime corresponds to $u\gg 1$, where the term quadratic
in the energy density dominates the Friedmann equation. Thus, in this limit,
\begin{equation}
H^2\simeq\frac{V^2}{6\mpl^2\lam}.
\label{eq:HEF}
\end{equation}
We mention that the numerical threshold used to identify this regime is model
dependent.
The condition $V/\lam>100$, adopted in the original analysis, corresponds
to $u>50$ and provides a useful phenomenological benchmark
\cite{LinNgCheung2019}.

Moreover, we use the reduced four-dimensional Planck mass $\mpl$, recalling that
the
relation to the five-dimensional Planck mass depends on convention. In the convention
of the original source, $\mplfour$ denotes the unreduced four-dimensional
Planck mass, with $\mpl=\mplfour/\sqrt{8\pi}$, and
\begin{equation}
\mplfour=\sqrt{\frac{3}{4\pi}}\,
\frac{\mplfive^3}{\sqrt{\lam}}.
\label{eq:PlanckRelation}
\end{equation}
Stating this convention explicitly is useful because equivalent RS-II
relations are often written in terms of reduced five-dimensional
parameters. The subsequent results are independent of the convention,
provided that $\lam$ and the Planck masses are defined consistently.

As we observe, the modification in the Friedmann equation \eqref{eq:braneF} arises
from the brane embedding
and the associated junction conditions. The scalar kinetic term remains
canonical, therefore $V(\phi)$ is the potential of the canonically normalized
brane field and no additional field redefinition is required at the
background level. Lastly, note that the perturbation sector has a distinct
higher-dimensional
structure, which we discuss when examining tensor perturbations.

\subsection{Monomial slow-roll dynamics and scalar observables}

We consider the monomial potential
\begin{equation}
V(\phi)=A\phi^p,
\qquad A>0,
\qquad p>0,
\label{eq:mono}
\end{equation}
and we focus on the branch with $\phi>0$ during inflation. For integer $p$,
the potential has its usual continuation to negative field values. For
fractional $p$, the continuation depends on the chosen branch, while for
$0<p<1$ the potential is nonanalytic at the origin. The inflationary
trajectory considered here terminates before reaching that point. Therefore, a
reheating scenario based on oscillations about $\phi=0$  requires
additional specification.

The potential derivatives are
\begin{equation}
\frac{V'}{V}=\frac{p}{\phi},
\ \ \
\frac{V''}{V}=\frac{p(p-1)}{\phi^2},
\ \ \
\frac{V^{(3)}}{V}
=\frac{p(p-1)(p-2)}{\phi^3},
\label{eq:derivativesmono1}
\end{equation}
and
\begin{equation}
\frac{V^{(4)}}{V}
=\frac{p(p-1)(p-2)(p-3)}{\phi^4}.
\label{eq:derivativesmono2}
\end{equation}
The higher derivatives will also enter the scalar weak gravity
diagnostics discussed in Sec. \ref{Swamplanddiagnostics}.

Now, for the brane background, the potential slow-roll parameters are
\begin{equation}
\epsilon_V
=\frac{M_{\mathrm{Pl}}^2}{2}
\left(\frac{V'}{V}\right)^2
\frac{1+V/\lambda}
{\left(1+V/(2\lambda)\right)^2},
\label{eq:epsbrane}
\end{equation}
and
\begin{equation}
\eta_V
=M_{\mathrm{Pl}}^2\frac{V''}{V}
\frac{1}{1+V/(2\lambda)}.
\label{eq:etabrane}
\end{equation}
In the high-energy limit, these expressions reduce to
\begin{equation}
\epsilon_V\simeq
2M_{\mathrm{Pl}}^2\lambda\frac{(V')^2}{V^3},
\qquad
\eta_V\simeq
2M_{\mathrm{Pl}}^2\lambda\frac{V''}{V^2}.
\label{eq:epsetaHE}
\end{equation}
Hence, for the potential in Eq.~\eqref{eq:mono}, this gives
\begin{eqnarray}
\label{eq:epsetaMono0}
&&\epsilon_V
=\frac{2p^2M_{\mathrm{Pl}}^2\lambda}
{A\phi^{p+2}},
\\
&&\eta_V
=\frac{2p(p-1)M_{\mathrm{Pl}}^2\lambda}
{A\phi^{p+2}},
\\
&&\frac{\eta_V}{\epsilon_V}=\frac{p-1}{p}.
\label{eq:epsetaMono}
\end{eqnarray}
Inflation ends when the largest relevant slow-roll parameter reaches
unity. For $p>\frac12$, one has $\epsilon_V>|\eta_V|$, whereas for
$0<p<\frac12$, the absolute value of the curvature parameter is larger.
The end-of-inflation field value is therefore
\begin{equation}
\phi_{\rm end}^{p+2}
=
\begin{cases}
\displaystyle
\frac{2p^2M_{\mathrm{Pl}}^2\lambda}{A},
& p\geq\frac12,\\[1.0ex]
\displaystyle
\frac{2p(1-p)M_{\mathrm{Pl}}^2\lambda}{A},
& 0<p<\frac12.
\end{cases}
\label{eq:phiend}
\end{equation}

The number of e-folds between horizon exit and the end of inflation is
\begin{equation}
N=\frac{1}{M_{\mathrm{Pl}}^2}
\int_{\phi_{\rm end}}^{\phi_*}
\frac{V}{V'}
\left(1+\frac{V}{2\lambda}\right)
\mathrm{d}\phi.
\label{eq:Ngeneral}
\end{equation}
and in the high-energy limit this becomes
\begin{equation}
N\simeq\frac{1}{2M_{\mathrm{Pl}}^2\lambda}
\int_{\phi_{\rm end}}^{\phi_*}
\frac{V^2}{V'}\,\mathrm{d}\phi.
\label{eq:NHE}
\end{equation}
For the monomial potential, the integral gives
\begin{equation}
N=
\frac{A}
{2pM_{\mathrm{Pl}}^2\lambda(p+2)}
\left(\phi_*^{p+2}-\phi_{\rm end}^{p+2}\right).
\label{eq:NmonoExact}
\end{equation}
For $N\simeq 50$--$60$, the contribution from $\phi_{\rm end}$ is
numerically small over most of the small-$p$ range considered here.
Accordingly,
\begin{equation}
\phi_*^{p+2}
\simeq
\frac{2Np(p+2)M_{\mathrm{Pl}}^2\lambda}{A}.
\label{eq:phiStar}
\end{equation}
Finally, substituting this expression into
\eqref{eq:epsetaMono0}-\eqref{eq:epsetaMono} yields
\begin{equation}
\epsilon_*\simeq\frac{p}{N(p+2)},
\qquad
\eta_*\simeq\frac{p-1}{N(p+2)}.
\label{eq:epsetaN}
\end{equation}

At leading order in slow roll, the scalar spectral index is
\begin{equation}
n_s-1\simeq 2\eta_*-6\epsilon_*
\simeq -\frac{2+4p}{N(p+2)},
\label{eq:nsmono}
\end{equation}
and the running is
\begin{equation}
\alpha_s\simeq
-\frac{2+4p}{N^2(p+2)}.
\label{eq:running}
\end{equation}
For the parameter range of interest, the running is of order
$10^{-4}$--$10^{-3}$ and remains small compared with the current
observational uncertainty. We mention that this conclusion follows within the
leading-order slow-roll treatment, and a more precise analysis would require
higher-order corrections and an exact mapping between the pivot scale
and the end of inflation.

The scalar curvature power spectrum in the slow-roll approximation is
\begin{equation}
A_s\simeq
\frac{1}{12\pi^2}
\frac{V^3}{M_{\mathrm{Pl}}^6V'^2}
\left(1+\frac{V}{2\lambda}\right)^3,
\label{eq:Asfull}
\end{equation}
and in the high-energy limit it becomes
\begin{equation}
A_s\simeq
\frac{V^6}
{96\pi^2M_{\mathrm{Pl}}^6\lambda^3V'^2}.
\label{eq:AsHE}
\end{equation}
For  the power-law potential (\ref{eq:mono}), relation \eqref{eq:phiStar} allows this
normalization to
be written as
\begin{equation}
A_s\simeq
\frac{\left[2Np(p+2)\right]^4}{96\pi^2p^2}
\left(\frac{\lambda}{M_{\mathrm{Pl}}^4}\right)
\left(\frac{M_{\mathrm{Pl}}}{\phi_*}\right)^6,
\label{eq:AsSimplified}
\end{equation}
which yields
\begin{equation}
\frac{\phi_*}{M_{\mathrm{Pl}}}
\simeq
\left[
\frac{\left(2Np(p+2)\right)^4}
{96\pi^2p^2A_s}
\right]^{1/6}
\left(\frac{\lambda}{M_{\mathrm{Pl}}^4}\right)^{1/6}.
\label{eq:phiAs}
\end{equation}
Note that  $A_s$ is evaluated at the pivot scale. Equation~\eqref{eq:phiAs}
shows how the scalar-amplitude normalization relates the pivot field
value to the brane tension. In particular, decreasing $\lambda$ decreases
$\phi_*$ while the normalization can be maintained by a compensating
change in $A$.

\subsection{Tensor perturbations and brane-specific consistency relations}

Concerning the tensor sector  one should be careful since tensor perturbations
propagate in the five-dimensional bulk. The scalar spectrum at leading
slow-roll order is given by Eq.~\eqref{eq:Asfull}, whereas the tensor
spectrum includes a bulk normalization factor. In the RS-II scenario it
can be written as
\begin{equation}
\mathcal{P}_T
=\frac{2H^2}{\pi^2 M_{\mathrm{Pl}}^2}F^2(z),
\qquad
z\equiv\frac{H}{\mu_{\mathrm{b}}},
\qquad
\mu_{\mathrm{b}}=\frac{1}{M_{\mathrm{Pl}}}\sqrt{\frac{\lambda}{6}},
\label{eq:PTbrane}
\end{equation}
where
$F^{-2}(z)=\sqrt{1+z^2}-z^2\sinh^{-1}(1/z)$ accounts for the normalization
of the tensor zero mode in the bulk, and $\mu_{\mathrm{b}}$ is the energy
scale set by the AdS curvature and the brane tension
\cite{KoyamaEtAl2007,CardosoEtAl2007}. In the high-energy regime
$z\simeq V/\lambda\gg1$ and $F^2\simeq 3z/2$, so that
$\mathcal{P}_T\simeq V^3/(2\pi^2M_{\mathrm{Pl}}^4\lambda^2)$.

As it is known, the
tensor-to-scalar ratio is
defined as $r\equiv\mathcal{P}_T/A_s$. Dividing the high-energy tensor spectrum by
Eq.~\eqref{eq:AsHE} gives $r\simeq48M_{\mathrm{Pl}}^2\lambda V'^2/V^3$,
which for the monomial model reads
\begin{equation}
r\simeq24\epsilon_*
\simeq\frac{24p}{N(p+2)}.
\label{eq:rmono}
\end{equation}
This relation applies within the same leading-order, high-energy
approximation used above. Obviously it should not be combined with the standard
four-dimensional relation $r=16\epsilon_V$.

Finally, combining Eq.~\eqref{eq:rmono} with the scalar spectral index in
Eq.~\eqref{eq:nsmono} gives
\begin{equation}
n_s\simeq1-\frac{1}{N}-\frac{r}{8}.
\label{eq:nsrNidentity}
\end{equation}
This is a leading-order consistency relation for high-energy monomial
inflation on the RS-II brane. It is independent of the potential
normalization $A$ and the brane tension $\lambda$, provided that the
high-energy approximation and the slow-roll treatment remain valid.

\section{Swampland diagnostics for the monomial model}
\label{Swamplanddiagnostics}

The Swampland criteria address different aspects of the possible
consistency of low-energy theories with quantum gravity
\cite{k,o,bbb,50,51}. We consider the de Sitter criteria, the scalar weak
gravity conjectures, the field-distance conjecture, and the
Trans-Planckian Censorship Conjecture. These conditions are conceptually
distinct and will be assessed separately. Their satisfaction will be treated
  as a diagnostic of the effective model, rather than as a
demonstration of a complete ultraviolet construction.

The background considered in this work is five-dimensional, while the
inflaton is a canonically normalized scalar field localized on the
four-dimensional brane. Therefore, we evaluate the potential-based
criteria using the four-dimensional effective field space of that
scalar. A genuinely five-dimensional Swampland bound would require a
specified five-dimensional scalar sector and its canonical
normalization. Since the model does not provide these ingredients or a
complete compactification, the results below should be understood as
four-dimensional effective-field-space diagnostics within an RS-II
background.

For a positive potential and a single canonical scalar, we define the
dimensionless slope and curvature parameters as
\begin{equation}
F_1\equiv\frac{M_{\mathrm{Pl}}|V'|}{V},
\qquad
F_2\equiv\frac{M_{\mathrm{Pl}}^2V''}{V}.
\label{eq:F1F2}
\end{equation}
Their relation to the inflationary observables differs from that in
four-dimensional Einstein gravity. In the high-energy RS-II regime,
Eqs.~\eqref{eq:epsetaHE} and \eqref{eq:rmono} give respectively
\begin{equation}
F_{1*}^2
\simeq u_*\epsilon_*
\simeq \frac{u_*r}{24},
\qquad
u_*\equiv\frac{V_*}{2\lambda},
\label{eq:slopeRelationBrane}
\end{equation}
and
\begin{equation}
F_{2*}
\simeq u_*\eta_*
\simeq u_*\left(\frac{n_s-1}{2}+\frac{r}{8}\right).
\label{eq:F2braneObs}
\end{equation}
Hence, the four-dimensional relation $F_1=\sqrt{r/8}$ does not apply in
the high-energy brane regime. Thus, the tensor-to-scalar ratio alone does not
fix the potential slope, and the brane parameter $u_*$ must also be
specified.

\begin{figure}[htbp]
  \centering
  \includegraphics[width=8.4cm]{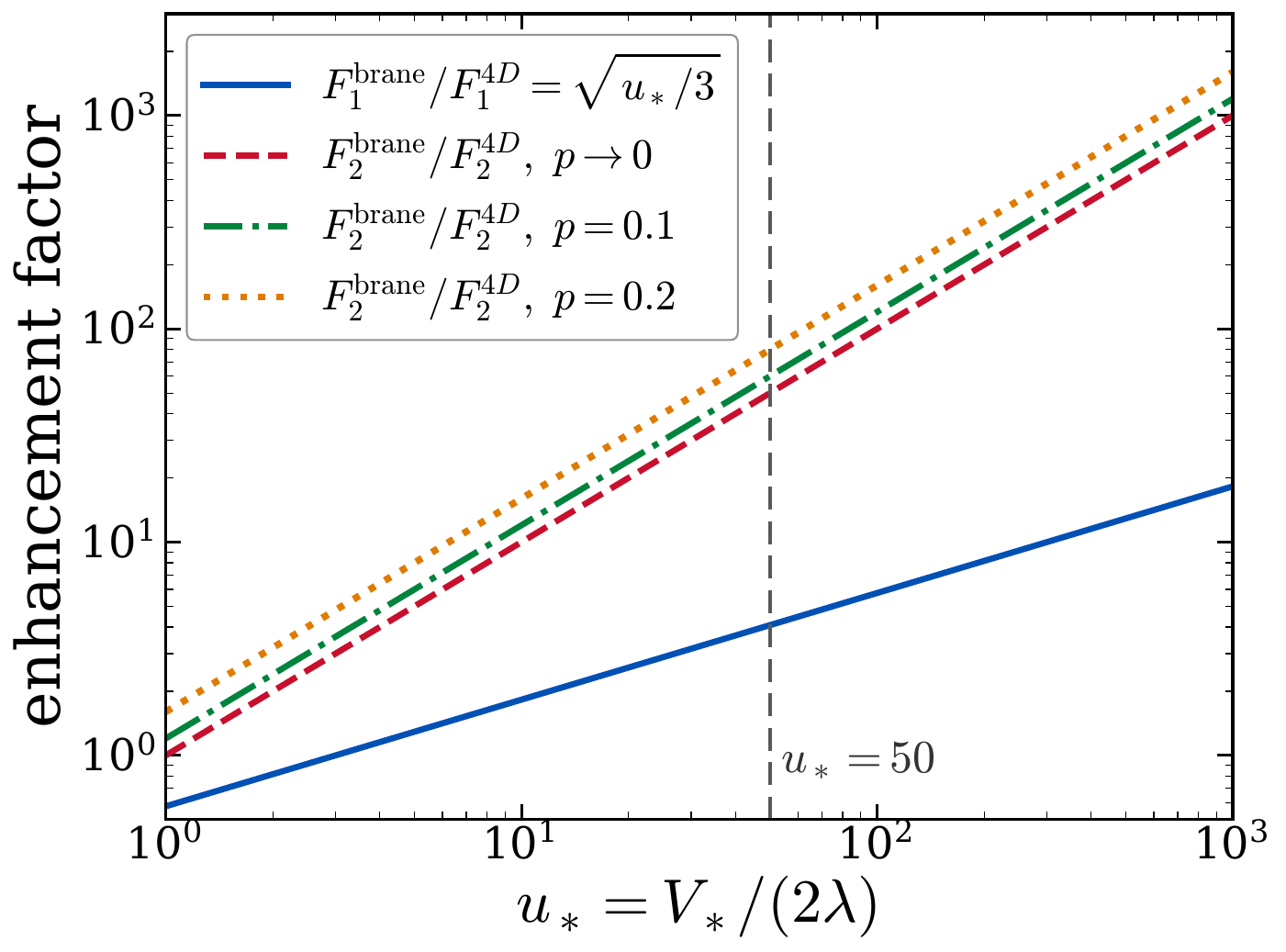}
 \caption{{\it{Ratios of the brane potential diagnostics to the values that the
four-dimensional slow-roll relations would assign to the same $(n_s,r)$,
Eq.~(\ref{eq:enhancement}), as functions of $u_*=V_*/(2\lambda)$. The slope ratio
$\sqrt{u_*/3}$ does not depend on the monomial power, while the curvature ratio
$u_*(1-p)/(1-5p/2)$ is drawn for $p\to0$, $p=0.1$ and $p=0.2$. The vertical dashed
line marks $u_*=50$, i.e. the benchmark $V_*/\lambda=100$ that identifies the
high-energy regime.}}}
  \label{fig4}
\end{figure}

In order to see these more transparently, in Fig. \ref{fig4} we  display the
enhancement of the brane potential diagnostics
relative to their four-dimensional counterparts. The four-dimensional relations
$r=16\epsilon_V$ and
$n_s-1=2\eta_V-6\epsilon_V$ would give $F_1^{4D}=\sqrt{r/8}$ and
$F_2^{4D}=(n_s-1)/2+3r/16$ for the same observables, so that
\begin{equation}
\frac{F_1^{\rm brane}}{F_1^{4D}}=\sqrt{\frac{u_*}{3}},
\qquad
\frac{F_2^{\rm brane}}{F_2^{4D}}
=u_*\,\frac{1-p}{1-5p/2},
\label{eq:enhancement}
\end{equation}
where the second ratio uses Eqs.~\eqref{eq:nsmono} and \eqref{eq:rmono}
and holds for $p<2/5$. Note that at $p=2/5$ the four-dimensional estimate of the
curvature vanishes. At the benchmark value
$u_*=50$, the ratio for the slope diagnostic is
$\sqrt{50/3}\simeq4.1$, while that for the curvature diagnostic is $50$
for $p\to0$ and $68$ for $p=0.15$. Therefore, the difference is  substantial even at
the adopted high-energy
benchmark, and the four-dimensional relations cannot be used directly to
infer the potential derivatives in the RS-II regime.

\subsection{de Sitter, refined de Sitter, and further-refined de Sitter criteria}
\label{IIIA}

The original de Sitter criterion requires the potential to have a
sufficiently steep slope, namely
\begin{equation}
F_1\geq c,
\qquad
c=\mathcal{O}(1).
\label{eq:DSSC}
\end{equation}
The refined de Sitter criterion allows this slope condition to be
replaced by sufficiently negative curvature. Therefore, for a single scalar it
 requires either an order-one slope or a sufficiently negative
value of $F_2$. The further-refined de Sitter criterion combines the
slope and curvature contributions and can be written as
\begin{equation}
F_1^q-\alpha F_2\geq1-\alpha,
\qquad
q>2,
\qquad
0<\alpha<1.
\label{eq:FRDSSC}
\end{equation}
The parameters $c$, $q$, and $\alpha$ are conjectural coefficients, i.e. they are
not parameters of the brane model and are not determined by
CMB observations.

For the monomial potential in Eq.~\eqref{eq:mono}, we introduce
$x\equiv\phi/M_{\mathrm{Pl}}$. The slope and curvature parameters then
take the form
\begin{equation}
F_1=\frac{p}{|x|},
\qquad
F_2=\frac{p(p-1)}{x^2}.
\label{eq:Fmono}
\end{equation}
The original de Sitter criterion consequently restricts the field value
to a finite range. The refined criterion can also be satisfied in the
$0<p<1$ branch, where $F_2<0$ and the curvature contribution is
negative. This possibility changes the relative importance of the slope
and curvature conditions, but does not remove their dependence on the
field value.

For the monomial potential, the further-refined criterion becomes
\begin{equation}
\left(\frac{p}{|x|}\right)^q
-\alpha\frac{p(p-1)}{x^2}
\geq1-\alpha.
\label{eq:FRmonomial}
\end{equation}
When $F_2<1$  this inequality gives the lower bound
\begin{equation}
\alpha\geq\alpha_{\min}
\equiv\frac{1-F_1^q}{1-F_2},
\label{eq:amin}
\end{equation}
while for $F_2>1$  it instead gives the upper bound
\begin{equation}
\alpha\leq\frac{F_1^q-1}{F_2-1}.
\label{eq:amax}
\end{equation}
These bounds must be considered together with $0<\alpha<1$.

In the $0<p<1$ branch, $F_2<0$, so the curvature term contributes
positively to Eq.~\eqref{eq:FRmonomial}. If $p/|x|\geq1$, the slope
term alone is sufficient to satisfy the further-refined criterion.
For smaller slopes, the negative curvature can still allow the
inequality to hold. This makes the small-$p$ branch favored by the
tensor constraint a natural region in which to examine the criterion.

For $1<p<2$, one has $F_2>0$, and satisfaction depends on the
competition between the slope and curvature terms. At sufficiently
small field values, the slope contribution grows more rapidly because
$q>2$. The criterion can therefore be satisfied in this regime as
well, although there is no parameter-independent satisfaction across
the full field range.

These results must be interpreted using the brane relations in
Eqs.~\eqref{eq:slopeRelationBrane} and \eqref{eq:F2braneObs}. In
particular, substituting an observed value of $r$ into the
four-dimensional relation $F_1=\sqrt{r/8}$ would give an incorrect
slope for the high-energy RS-II background. Mapping the
further-refined criterion to observations requires specifying $u_*$,
or relating it to the scalar-amplitude normalization and the brane
tension.

\subsection{Scalar weak gravity conjectures}
\label{IIIB}

The scalar weak gravity conjecture (SWGC) and its strong version
(SSWGC) constrain higher derivatives of the scalar potential. In the
canonical single-field convention of Gonzalo and Ib\'a\~nez
\cite{GonzaloIbanez2019}, the two conditions are
\begin{equation}
M_{\mathrm{Pl}}^2\bigl(V^{(3)}\bigr)^2
\geq \bigl(V''\bigr)^2,
\label{eq:SWGCdimension}
\end{equation}
and
\begin{equation}
2M_{\mathrm{Pl}}^2\bigl(V^{(3)}\bigr)^2
-M_{\mathrm{Pl}}^2V''V^{(4)}
\geq \bigl(V''\bigr)^2,
\label{eq:SSWGCdimension}
\end{equation}
respectively. We define the dimensionless derivative combinations
\begin{equation}
d_k\equiv\frac{M_{\mathrm{Pl}}^k V^{(k)}}{V},
\label{eq:dk}
\end{equation}
and therefore  the conjectures become
\begin{equation}
\Delta_{\mathrm{SWGC}}\equiv d_3^2-d_2^2\geq0,
\label{eq:DeltaSW}
\end{equation}
and
\begin{equation}
\Delta_{\mathrm{SSWGC}}\equiv
2d_3^2-d_2d_4-d_2^2\geq0.
\label{eq:DeltaSSW}
\end{equation}

The application of these conjectures as off-shell inequalities over an
entire field range remains debated. Some analyses instead relate them
to scattering amplitudes or to specific points in field space
\cite{BenakliBranchinaLafforgue2020,DudasEtAl2023}. We therefore treat
them as diagnostic conditions rather than experimentally established
requirements.

Using Eq.~\eqref{eq:derivativesmono1} and
$x\equiv\phi/M_{\mathrm{Pl}}$, the monomial potential gives
\begin{equation}
d_2=\frac{p(p-1)}{x^2},
\qquad
d_3=\frac{p(p-1)(p-2)}{x^3},
\end{equation}
and
\begin{equation}
d_4=\frac{p(p-1)(p-2)(p-3)}{x^4}.
\end{equation}
The SWGC diagnostic is then
\begin{equation}
\Delta_{\mathrm{SWGC}}
=\frac{p^2(p-1)^2}{x^6}
\left[(p-2)^2-x^2\right].
\label{eq:SWmono}
\end{equation}
Away from the degenerate points, the condition
$\Delta_{\mathrm{SWGC}}\geq0$ requires
\begin{equation}
|x|\leq|p-2|,
\label{eq:SWwindow}
\end{equation}
and thus for $0<p<2$ and $p\neq1$  this gives the pivot-field bound
\begin{equation}
|\phi_*|\leq(2-p)M_{\mathrm{Pl}}.
\label{eq:SWwindowphi}
\end{equation}
At $p=1$, the derivative combinations entering the SWGC vanish and
the diagnostic is saturated identically, while at $p=2$  the formal upper
limit is zero, which does not give a finite nonzero field value at
which the normalized derivative condition is satisfied.

For the SSWGC, substitution of the same derivatives gives
\begin{equation}
\Delta_{\mathrm{SSWGC}}
=\frac{p^2(p-1)^2}{x^6}
\left[(p-1)(p-2)-x^2\right].
\label{eq:SSWmono}
\end{equation}
Away from the degenerate points, the condition requires
\begin{equation}
x^2\leq(p-1)(p-2).
\label{eq:SSWwindowGeneral}
\end{equation}
Therefore, the allowed branches are
\begin{equation}
\begin{array}{ll}
0<p<1:
&\displaystyle |x|\leq\sqrt{(1-p)(2-p)},\\[1.0ex]
p>2:
&\displaystyle |x|\leq\sqrt{(p-1)(p-2)}.
\end{array}
\label{eq:SSWbranches}
\end{equation}
For $1<p<2$, the right-hand side of
Eq.~\eqref{eq:SSWwindowGeneral} is negative, and thus  there is no finite
nonzero solution, while at $p=1$ we have $d_2=d_3=d_4=0$ and the SSWGC diagnostic
vanishes identically.

In the phenomenologically relevant $0<p<1$ branch, the SSWGC permits
a narrower field interval than the SWGC. For example, at $p=1/2$,
the SSWGC gives $|\phi_*|\leq\sqrt{3/4}\,M_{\mathrm{Pl}}$, whereas
the SWGC gives $|\phi_*|\leq1.5M_{\mathrm{Pl}}$. The scalar-amplitude
normalization in Eq.~\eqref{eq:phiAs} gives
$\phi_*\propto\lambda^{1/6}$. Hence, decreasing  the brane tension can
place the pivot field value within these finite intervals, provided
that the resulting parameters remain in the high-energy regime and
within the validity range of the five-dimensional effective
description.

We mention that neither condition is satisfied over the full field domain. In
particular, both diagnostics become negative at sufficiently large
$|x|$. Thus, the results  identify finite field intervals in which
the conjectural conditions hold. They do not establish global
satisfaction along the inflationary trajectory or demonstrate that
the effective theory has a consistent quantum-gravity completion.

\subsection{Field distance and trans-Planckian censorship}
\label{IIIC}

Since the inflaton has a canonical kinetic term, the proper
one-dimensional field-space distance between horizon exit and the end
of inflation is
\begin{equation}
\Delta\phi_{\rm proper}
=|\phi_*-\phi_{\rm end}|.
\end{equation}
For the rolling branch considered here, $\phi_*>\phi_{\rm end}$, and
we can write
\begin{equation}
\Delta\phi=\phi_*-\phi_{\rm end}.
\label{eq:Deltaphi}
\end{equation}
Now, a conservative field-range version of the distance criterion is
\begin{equation}
\frac{\Delta\phi}{M_{\mathrm{Pl}}}
\lesssim c_{\rm dist},
\qquad
c_{\rm dist}=\mathcal{O}(1).
\label{eq:SDC}
\end{equation}
Using Eqs.~\eqref{eq:phiend} and \eqref{eq:phiStar}, the ratio of the
field value at the end of inflation to its value at horizon exit is
\begin{equation}
\frac{\phi_{\rm end}}{\phi_*}
\simeq
\begin{cases}
\displaystyle
\left[\frac{p}{N(p+2)}\right]^{1/(p+2)},
& p\geq\frac12,\\[1.2ex]
\displaystyle
\left[\frac{1-p}{N(p+2)}\right]^{1/(p+2)},
& 0<p<\frac12.
\end{cases}
\label{eq:phiendratio}
\end{equation}
Hence, it follows that
\begin{equation}
\frac{\Delta\phi}{M_{\mathrm{Pl}}}
\simeq
\frac{\phi_*}{M_{\mathrm{Pl}}}
\left(1-\frac{\phi_{\rm end}}{\phi_*}\right).
\label{eq:Deltax}
\end{equation}
Combining this result with the scalar-amplitude normalization in
 \eqref{eq:phiAs}, a sufficient condition for a sub-Planckian
excursion is
\begin{equation}
\left[
\frac{\left(2Np(p+2)\right)^4}
{96\pi^2p^2A_s}
\right]^{1/6}
\left(\frac{\lambda}{M_{\mathrm{Pl}}^4}\right)^{1/6}
\left(1-\frac{\phi_{\rm end}}{\phi_*}\right)
\lesssim c_{\rm dist}.
\label{eq:SDClambda}
\end{equation}
This condition can be expressed as an upper bound on the dimensionless
brane tension, namely
\begin{eqnarray}
&&\frac{\lambda}{M_{\mathrm{Pl}}^4}
\lesssim
\frac{c_{\rm dist}^{\,6}}{\mathcal{K}(p,N)}
\left(1-\frac{\phi_{\rm end}}{\phi_*}\right)^{-6},
\end{eqnarray}
with
\begin{eqnarray}
&&
\mathcal{K}(p,N)\equiv
\frac{\left[2Np(p+2)\right]^4}{96\pi^2p^2A_s}.
\label{eq:SDClambdaExplicit}
\end{eqnarray}

On the other hand, the modern distance conjecture is stronger than a bound on the
field
excursion. It associates a parametrically large geodesic distance
with an infinite tower of states whose masses decrease exponentially,
namely
\begin{equation}
m_{\rm tower}\sim
m_0\exp\left(
-\alpha_{\rm dist}\frac{\Delta\phi}{M_{\mathrm{Pl}}}
\right),
\qquad
\alpha_{\rm dist}=\mathcal{O}(1).
\label{eq:tower}
\end{equation}
The present brane model does not specify such a tower. Although the
five-dimensional setup contains bulk gravitational modes, their
spectrum and dependence on the inflaton are not specified. The
trajectory can therefore be made sub-Planckian, so it need not violate
the field-range version of the distance criterion. However, the full distance
conjecture cannot be declared satisfied or violated without
an explicit tower construction.

For fractional powers with $0<p<1$, higher derivatives of the
potential become singular as $\phi\to0^+$. Thus, the scalar derivative
criteria discussed in Sec. \ref{IIIB} should  be applied only
within a controlled effective-field-theory domain.

The trans-Planckian censorship conjecture (TCC) requires that quantum
fluctuations whose physical wavelengths begin below the Planck scale
do not cross the Hubble radius during a controlled accelerating phase.
In its simplest quasi-de Sitter form, this condition is
\begin{equation}
e^{N_{\rm TCC}}H_{\rm e}\lesssim M_{\mathrm{Pl}},
\label{eq:TCCbasic}
\end{equation}
where $N_{\rm TCC}$ is the number of e-folds relevant to the TCC bound
and $H_{\rm e}$ is the Hubble parameter at the end of the accelerating
phase. If the variation of $H$ over this interval is neglected, one
may replace $H_{\rm e}$ by $H_*$. The quantity $N_{\rm TCC}$ is not
necessarily identical to the pivot-to-end e-fold number $N_*$ used in
the inflationary observables \cite{BedroyaVafa2019,BedroyaEtAl2019}.

In the high-energy brane regime, the relation between the scalar
amplitude, the Hubble parameter, the slow-roll parameter, and the
brane tension differs from its four-dimensional counterpart. Using
expression \eqref{eq:AsHE} together with Eq.~\eqref{eq:HEF}, we obtain
\begin{equation}
A_s\simeq
\frac{\sqrt{6}}{8\pi^2}
\frac{H_*^3}
{M_{\mathrm{Pl}}\sqrt{\lambda}\,\epsilon_*}.
\label{eq:AsHbrane}
\end{equation}
With $r\simeq24\epsilon_*$, this gives
\begin{equation}
r\simeq
\frac{3\sqrt{6}}{\pi^2A_s}
\frac{H_*^3}
{M_{\mathrm{Pl}}\sqrt{\lambda}}.
\label{eq:rHlambda}
\end{equation}
Applying Eq.~\eqref{eq:TCCbasic} in the constant-$H$ approximation,
with $H_*$ replacing $H_{\rm e}$, yields the conditional bound
\begin{equation}
r\lesssim
\frac{3\sqrt{6}}{\pi^2A_s}
\frac{M_{\mathrm{Pl}}^2}{\sqrt{\lambda}}
e^{-3N_{\rm TCC}}.
\label{eq:rTCCbrane}
\end{equation}
Equivalently, the TCC gives an upper bound on the brane tension, i.e.
\begin{equation}
\frac{\lambda}{M_{\mathrm{Pl}}^4}
\lesssim
\frac{3}{32\pi^4A_s^2\epsilon_*^2}
e^{-6N_{\rm TCC}}
\simeq
\frac{54}{\pi^4A_s^2r^2}
e^{-6N_{\rm TCC}}.
\label{eq:lambdaTCC}
\end{equation}
In the monomial model $r$ is not free, i.e. it is fixed by $p$ and $N$ through
Eq.~\eqref{eq:rmono}, thus Eq.~\eqref{eq:lambdaTCC} is a bound on the brane
tension alone. Since the accelerated phase lasts at least as long as the
observable e-folds, $N_{\rm TCC}\geq N_*$. For $p=0.15$ and
$N_{\rm TCC}=60$,  \eqref{eq:lambdaTCC} gives
$\lambda/M_{\mathrm{Pl}}^4\lesssim10^{-136}$, and about $10^{-110}$ for
$N_{\rm TCC}=50$. Both values lie far below the lower
bound $\lambda/M_{\mathrm{Pl}}^4\gtrsim5\times10^{-65}$ that follows from
the five-dimensional gravitational scale given in \eqref{eq:lambdaLowerM5}.
Read the other way, that lower bound allows the TCC only for
$N_{\rm TCC}\lesssim33$, with a weak, logarithmic dependence on $p$. On the other
hand, the
P-ACT-LB tilt interval of Sec. \ref{CMBconstraints} requires
$N_*\gtrsim31$ at leading order, so the two conditions overlap only in a
narrow corner, $N_*\simeq31$--$33$ and $p\lesssim0.04$, with $\lambda$
close to its lower limit and a low inflationary scale,
$V_*^{1/4}\sim10^{7}\,\mathrm{GeV}$. Whether so few e-folds can be
realized depends on the reheating history, i.e.  Eq.~\eqref{eq:Nreh}. In this
corner we have $r\lesssim0.015$, which is not negligible, namely  in four-dimensional
single-field inflation the TCC forces $r\lesssim10^{-30}$
\cite{BedroyaEtAl2019}, whereas on the brane $r$ is set by $p$ and $N$ and
not by $H_*/M_{\mathrm{Pl}}$. The benchmark values $N=50$--$70$ used in the
figures below do not satisfy the TCC. Related studies of TCC in brane
inflation likewise obtain a brane-tension constraint
\cite{MohammadiGolanbariNasriSaaidi2022}.

This statement uses the four-dimensional reduced Planck mass as the
ultraviolet scale. If the relevant scale is instead the five-dimensional
one, $M_5<M_{\mathrm{Pl}}$, the condition becomes stronger and the corner shrinks.
The TCC must also
be considered together with the high-energy assumption. From Eq.~\eqref{eq:HEF} we
have
\begin{equation}
\frac{V}{2\lambda}
=\sqrt{\frac{3}{2}}\,
\frac{M_{\mathrm{Pl}}H}{\sqrt{\lambda}}.
\label{eq:uHlambda}
\end{equation}
Therefore, maintaining $V/(2\lambda)\gg1$ while imposing
$H\lesssim M_{\mathrm{Pl}}e^{-N_{\rm TCC}}$  requires a
sufficiently small brane tension,
\begin{equation}
\sqrt{\lambda}
\ll
\sqrt{\frac{3}{2}}\,
M_{\mathrm{Pl}}^2e^{-N_{\rm TCC}}.
\label{eq:lambdaTCCHE}
\end{equation}
The high-energy regime and the TCC are consequently not independent
assumptions. In particular, equation \eqref{eq:lambdaTCCHE} is much weaker than
 \eqref{eq:lambdaTCC}, because it does not use the scalar normalization, hence
it is the normalization that pushes the TCC limit down to the edge of the
allowed range of $\lambda$.

 \section{CMB constraints and combined viability}
 \label{CMBconstraints}

We now compare the leading-order predictions of the high-energy
monomial model with CMB observations. Since the observational
constraints depend on the data combination and on the pivot scale,
these choices are stated explicitly. We first use the scalar spectral
index and tensor-to-scalar ratio to identify the allowed region in the
$(p,N)$ plane. The scalar-amplitude normalization and the intersection
with the Swampland conditions are considered in the following
subsections.

\subsection{CMB benchmarks and constraints on the monomial power and
e-fold number}
\label{CMBconstraintssec}

As we mentioned in the Introduction,
Planck 2018 gives
\begin{equation}
n_s=0.9649\pm0.0042
\end{equation}
at 68\% confidence level. Combining Planck with BICEP/Keck BK15 gives
the 95\% upper limit
\begin{equation}
r_{0.002}<0.056.
\end{equation}
Moreover, the BICEP/Keck analysis through the 2018 observing season gives the
independent 95\% bound
\begin{equation}
r_{0.05}<0.036.
\end{equation}
On the other hand,
ACT DR6 alone gives $n_s=0.9666\pm0.0077$. The Planck and ACT
combination gives $n_s=0.9709\pm0.0038$, while adding CMB lensing and
DESI Year 1 BAO data gives the higher-tilt P-ACT-LB result
\begin{equation}
n_s=0.9743\pm0.0034.
\end{equation}
We mention that these values correspond to different likelihood combinations and
should be interpreted separately. For comparison with both the scalar
tilt and the tensor amplitude, we also use the P-ACT-LB-BK18 benchmark
\begin{equation}
n_s=0.9743\pm0.0034,
\qquad
r_{0.05}<0.038,
\end{equation}
with the quoted confidence levels.

\begin{figure}[htbp]
  \centering
  \includegraphics[width=8.4cm]{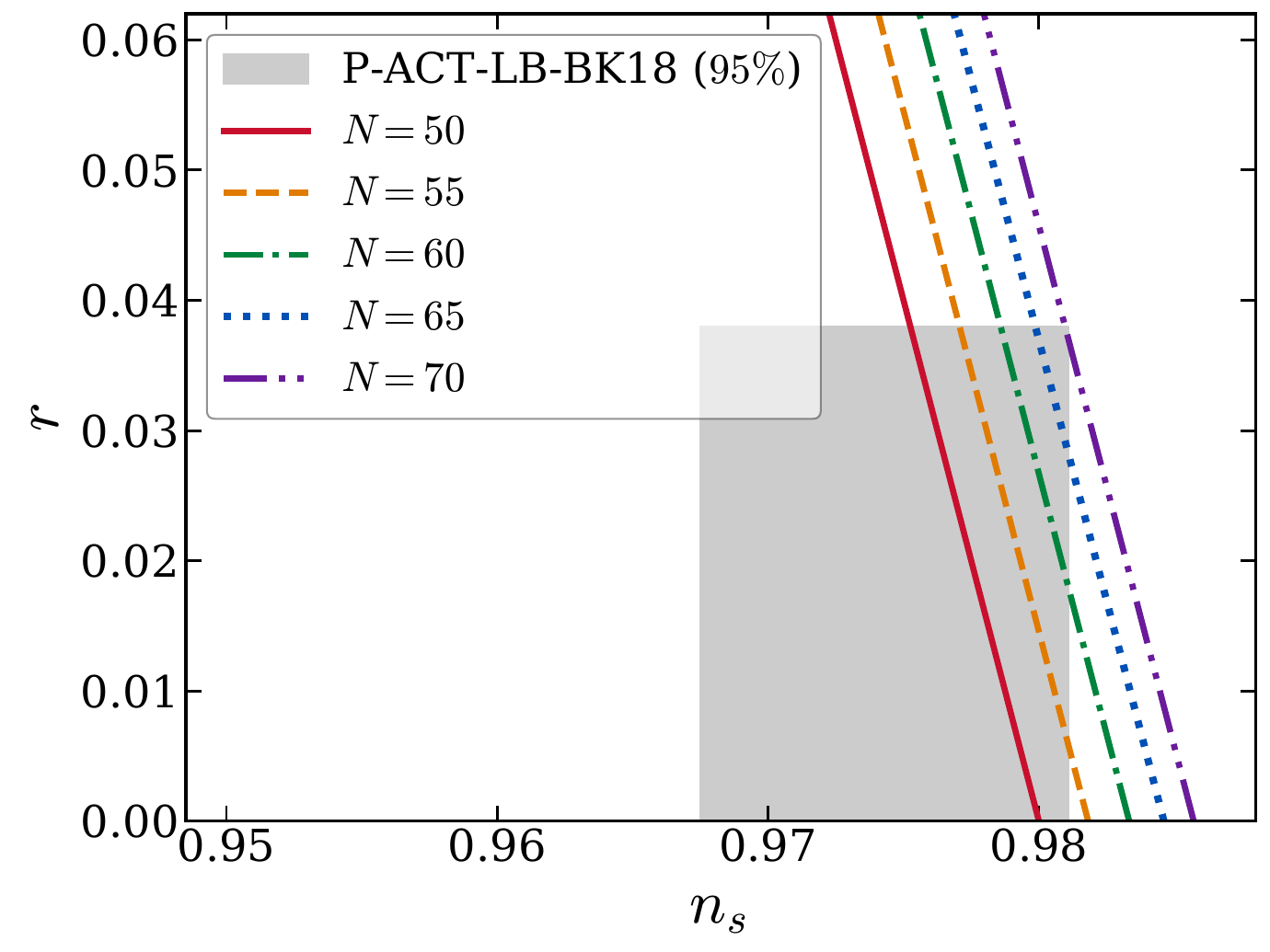}
 \caption{{\it{Leading-order predictions in the $(n_s,r)$ plane for high-energy
RS-II monomial inflation, for $N=50,55,60,65,$ and $70$, with the monomial power
running from $p=0$ on the horizontal axis upward. The shaded rectangle is the
$95\%$ P-ACT-LB-BK18 benchmark, $0.9675\lesssim n_s\lesssim0.9811$ and
$r_{0.05}<0.038$. Each line is the consistency relation $n_s\simeq1-1/N-r/8$ at
fixed $N$, hence a larger power gives a smaller $n_s$ and a larger $r$.}}}
  \label{fig1}
\end{figure}

In
Fig. \ref{fig1} we depict the model predictions in the $(n_s,r)$ plane against
the adopted P-ACT-LB-BK18 benchmark. At fixed $N$, increasing $p$ decreases
$n_s$ and raises $r$, while the predicted points satisfy the leading-order
consistency relation in Eq.~(\ref{eq:nsrNidentity}). The tensor bound can therefore
be
translated into an upper limit on $p$ for each chosen value of $N$.

The model predictions in Eqs.~\eqref{eq:nsrNidentity} and
\eqref{eq:rmono} allow the tensor bound to be expressed as a limit on
the monomial power. Solving the tensor relation for $p$ gives
\begin{equation}
p=\frac{2Nr}{24-Nr}.
\label{eq:pfromr}
\end{equation}
Hence, an upper bound on $r$  gives an upper bound on $p$, provided
that the model prediction and the observational bound refer to the
same pivot scale. For the P-ACT-LB-BK18 tensor bound, this yields
\begin{equation}
p<
\frac{0.076N}{24-0.038N}.
\label{eq:pmaxPACT}
\end{equation}
At the tensor boundary, the corresponding scalar tilt is
\begin{equation}
n_{s,\min}(N)
=1-\frac{1}{N}-\frac{0.038}{8},
\label{eq:nsminPACT}
\end{equation}
and at fixed $N$, decreasing $p$ decreases $r$ and increases the predicted
scalar tilt.

Finally, in Table~\ref{tab:PACTlimits} we provide
representative values at the tensor boundary, which   summarize the
leading-order
relations and do not replace a likelihood analysis. The P-ACT-LB scalar constraint
has an approximate 95\% interval
$0.9675\lesssim n_s\lesssim0.9811$. As we observe, the  values in
Table~\ref{tab:PACTlimits} show that the model enters this interval
over part of the displayed range in $N$. At $N=50$, the tilt at the
tensor boundary is above the central value but remains within the
quoted uncertainty. At larger $N$, the comparison becomes increasingly
sensitive to the upper edge of the interval and to the assumed
post-inflationary history.

\begin{table}[htbp]
\centering
\caption{Upper limits on the monomial power from
$r_{0.05}<0.038$ and the corresponding scalar tilt at the tensor
boundary.}
\label{tab:PACTlimits}
\begin{tabular}{ccc}
\hline
$N$ & $p_{\max}$ & $n_s(r_{0.05}=0.038)$ \\
\hline
50 & 0.1719 & 0.97525 \\
55 & 0.1908 & 0.97707 \\
60 & 0.2099 & 0.97858 \\
65 & 0.2294 & 0.97987 \\
70 & 0.2493 & 0.98096 \\
\hline
\end{tabular}
\end{table}

The Planck tensor bound is quoted at $k=0.002\,{\rm Mpc}^{-1}$,
whereas the scalar tilt and the BK18 bound are conventionally given
at $k_*=0.05\,{\rm Mpc}^{-1}$. In the high-energy RS-II limit we get
$n_t\simeq-r/8$. Thus, the corresponding leading-order pivot conversion is
\begin{equation}
r_{0.05}
\simeq r_{0.002}
\left(\frac{0.05}{0.002}\right)^{1-n_s-r/8}
\simeq r_{0.002}\,25^{1/N},
\label{eq:PlanckPivotConversion}
\end{equation}
where the final form follows from Eq.~\eqref{eq:nsrNidentity}. The
Planck limit then implies
\begin{equation}
r_{0.05}\lesssim0.056\,25^{1/N}
\end{equation}
and
\begin{equation}
p_{\max}^{\rm Planck}(N)
\simeq
\frac{0.112N\,25^{1/N}}
{24-0.056N\,25^{1/N}},
\label{eq:pmaxPlanck}
\end{equation}
and the minimum scalar tilt at the tensor boundary is
\begin{equation}
n_{s,\min}^{\rm Planck}(N)
\simeq
1-\frac{1}{N}
-\frac{0.056\,25^{1/N}}{8}.
\label{eq:nsminPlanck}
\end{equation}
Using the approximate 95\% upper edge $n_s\simeq0.9733$ of the quoted
Planck scalar constraint, the overlap extends only to about $N=52$
in this leading-order comparison. Therefore, the  Planck-compatible region is
 concentrated near the lower end of the considered
e-fold range.
\begin{figure}[htbp]
  \centering
  \includegraphics[width=8.4cm]{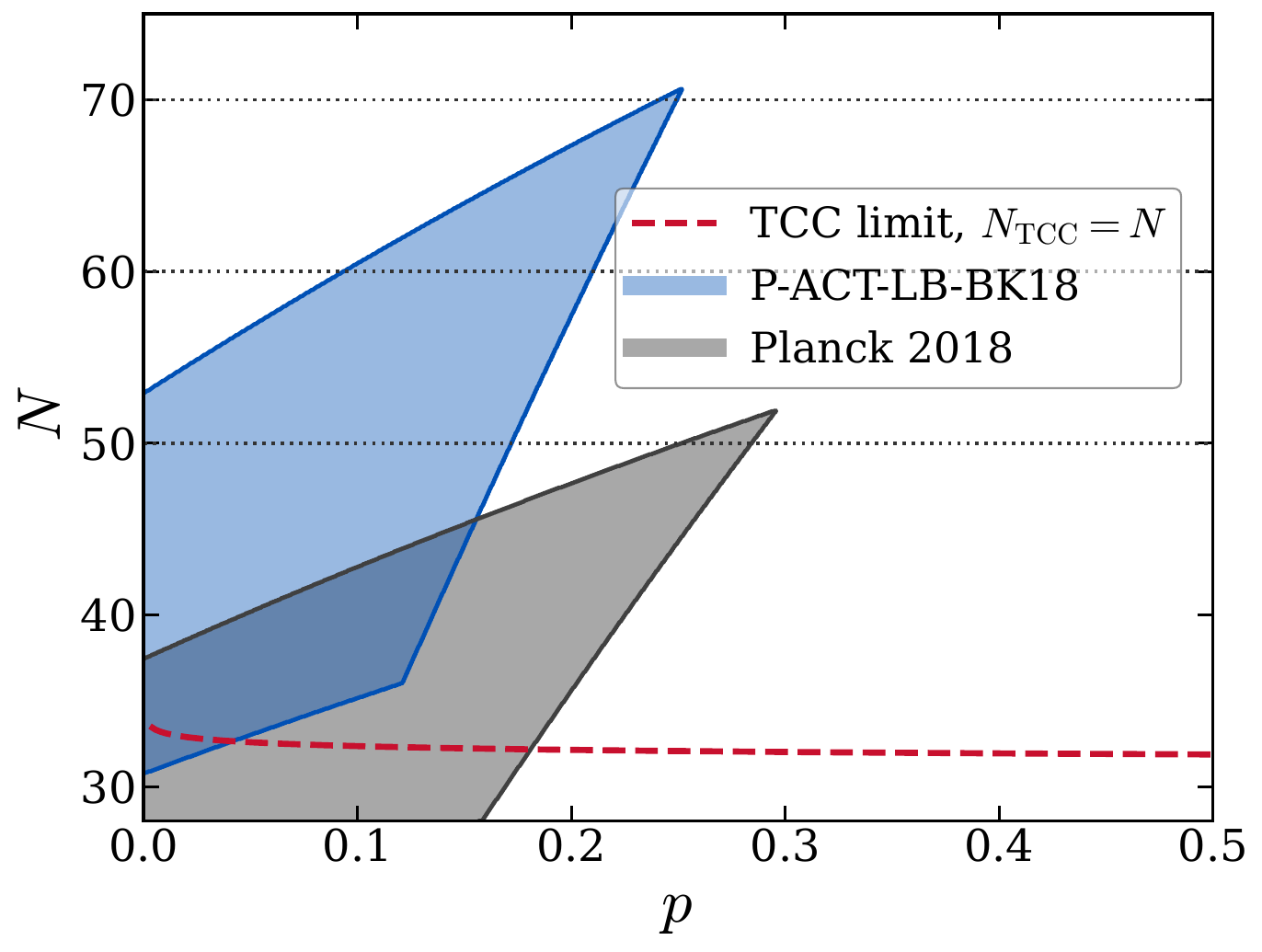}
\caption{{\it{Observational viability in the $(p,N)$ plane from the leading-order
high-energy RS-II predictions. The blue region satisfies the $95\%$
P-ACT-LB-BK18 scalar-tilt interval and the tensor bound $r_{0.05}<0.038$.
The gray region is the corresponding $95\%$ Planck 2018 region, with the
tensor bound converted to the $k_*=0.05\,{\rm Mpc}^{-1}$ pivot through
Eq.~(\ref{eq:PlanckPivotConversion}). The dotted lines mark $N=50$, $60$ and $70$.
Below the dashed curve the TCC limit of Eq.~(\ref{eq:lambdaTCC}) with
$N_{\rm TCC}=N$ is compatible with the lower bound on the brane tension,
Eq.~(\ref{eq:lambdaLowerM5}), and it meets the blue region only for
$N\simeq31$--$33$ and $p\lesssim0.04$.}}}
  \label{fig2}
\end{figure}

We mention that the value of $N$ is not directly measured, since it depends on the
post-inflationary equation of state, the reheating temperature, the
expansion history between the end of inflation and radiation
domination, and the pivot scale. A standard phenomenological
correction to the pivot-to-end e-fold relation is
\begin{equation}
\Delta N_{\rm reh}
=
\frac{1-3w_{\rm reh}}{12(1+w_{\rm reh})}
\ln\left(\frac{\rho_{\rm reh}}{\rho_{\rm end}}\right),
\label{eq:Nreh}
\end{equation}
where $w_{\rm reh}$ is the effective equation-of-state parameter
during reheating. In the simplest histories,
$-1/3<w_{\rm reh}<1$. Thus, the allowed regions in $(p,N)$ should be
read conditionally on the assumed reheating history. Choosing a single
value, such as $N=60$, provides a benchmark rather than a
reheating-independent prediction.

The reheating phase  also deserves particular attention for fractional powers
with $0<p<1$. Since the potential is nonanalytic at $\phi=0$, the
usual oscillatory phase around a quadratic minimum does not generally
follow from the monomial potential alone, and thus additional interactions or
a nonstandard reheating mechanism may be required. Therefore, the e-fold number
used in the CMB comparison is linked to the post-inflationary
dynamics.

The resulting allowed regions in the $(p,N)$ plane are shown in
Fig.~\ref{fig2}. As we observe, the P-ACT-LB-BK18 combination permits a broader
range of
sublinear powers, while the Planck-compatible region is concentrated toward
smaller values of $N$. The comparison uses the leading-order pivot conversion
in Eq.~(\ref{eq:PlanckPivotConversion}) and remains conditional on the assumed
post-inflationary history.

\subsection{Scalar normalization, field range, and brane-tension bounds}
\label{IVBnorm}

The scalar amplitude measured by CMB observations is
$A_s\simeq2.1\times10^{-9}$ at the pivot scale. Once this normalization
is imposed, Eq.~\eqref{eq:phiAs} relates the pivot field value to the
monomial power, the e-fold number, and the brane tension. In
particular, using the definition of $\mathcal{K}(p,N)$ in
(\ref{eq:SDClambdaExplicit}),
Eq.~\eqref{eq:phiAs} can be written as
\begin{equation}
x_*^6
=\mathcal{K}(p,N)\frac{\lambda}{M_{\mathrm{Pl}}^4},
\qquad
x_*\equiv\frac{\phi_*}{M_{\mathrm{Pl}}}.
\label{eq:xstarLambda}
\end{equation}
Thus, for fixed $p$ and $N$, decreasing  the brane tension decreases the
pivot field value. The dependence is proportional to
$\lambda^{1/6}$, therefore even a large change in $\lambda$ produces a more
modest change in $\phi_*$.

Hence, the field-space conditions derived in Sec.~\ref{Swamplanddiagnostics} can
be translated into upper bounds on the brane tension. For the
nondegenerate branches, the local bounds take the form
\begin{equation}
\frac{\lambda}{M_{\mathrm{Pl}}^4}
\lesssim
\frac{(p/c)^6}{\mathcal{K}(p,N)}
\qquad
\text{(de Sitter slope criterion)},
\label{eq:lambdaDSSC}
\end{equation}
\begin{equation}
\frac{\lambda}{M_{\mathrm{Pl}}^4}
\lesssim
\frac{(2-p)^6}{\mathcal{K}(p,N)}
\qquad
\text{(SWGC)},
\label{eq:lambdaSWGC}
\end{equation}
and
\begin{equation}
\frac{\lambda}{M_{\mathrm{Pl}}^4}
\lesssim
\frac{\left[(1-p)(2-p)\right]^3}{\mathcal{K}(p,N)}
\qquad
\text{(SSWGC, $0<p<1$)}.
\label{eq:lambdaSSWGC}
\end{equation}
The distance criterion constrains the total excursion rather than
$\phi_*$ alone. Using Eq.~\eqref{eq:Deltax}, its corresponding
sufficient bound is
\begin{equation}
\frac{\lambda}{M_{\mathrm{Pl}}^4}
\lesssim
\frac{c_{\rm dist}^{\,6}}
{\mathcal{K}(p,N)}
\left(1-\frac{\phi_{\rm end}}{\phi_*}\right)^{-6}.
\label{eq:lambdaSDC}
\end{equation}
The further-refined de Sitter condition depends also on the
conjectural parameters $\alpha$ and $q$, and on the brane quantity
$u_*$ defined in Sec. \ref{IIIA}.

For $0<p<1$ and $c=1$, the de Sitter slope criterion requires
$|x_*|\leq p$. Moreover, the SWGC allows the wider interval
$|x_*|\leq2-p$, while the SSWGC gives
$|x_*|\leq\sqrt{(1-p)(2-p)}$. The SSWGC bound is weaker than the
slope bound for $0<p<2/3$ and more restrictive for $2/3<p<1$.
For example, at $p=0.1$, the respective upper limits are
$|x_*|\leq0.1$, $|x_*|\leq1.9$, and
$|x_*|\leq\sqrt{1.71}$. Hence, the de Sitter slope condition is
the most restrictive of these three local conditions at this value
of $p$. The further-refined condition can be less restrictive in
this branch because $V''<0$.

The high-energy approximation imposes a further upper bound on the
brane tension, and  we require
\begin{equation}
\frac{V_{\rm end}}{\lambda}\gtrsim100.
\label{eq:HEendcriterion}
\end{equation}
Since the potential decreases during the rolling phase, the end of
inflation is the most restrictive point at which to impose this
condition. In terms of the end-field value, the slow-roll
expressions give
\begin{equation}
\frac{V_{\rm end}}{\lambda}
\simeq
\begin{cases}
\displaystyle
\frac{2p^2}{x_{\rm end}^2},
& p\geq\frac12,\\[1.0ex]
\displaystyle
\frac{2p(1-p)}{x_{\rm end}^2},
& 0<p<\frac12,
\end{cases}
\qquad
x_{\rm end}\equiv\frac{\phi_{\rm end}}{M_{\mathrm{Pl}}}.
\label{eq:VendLambda}
\end{equation}
Combining this requirement with the amplitude normalization gives
an additional upper limit on $\lambda$.

\begin{figure}[htbp]
  \centering
  \includegraphics[width=8.4cm]{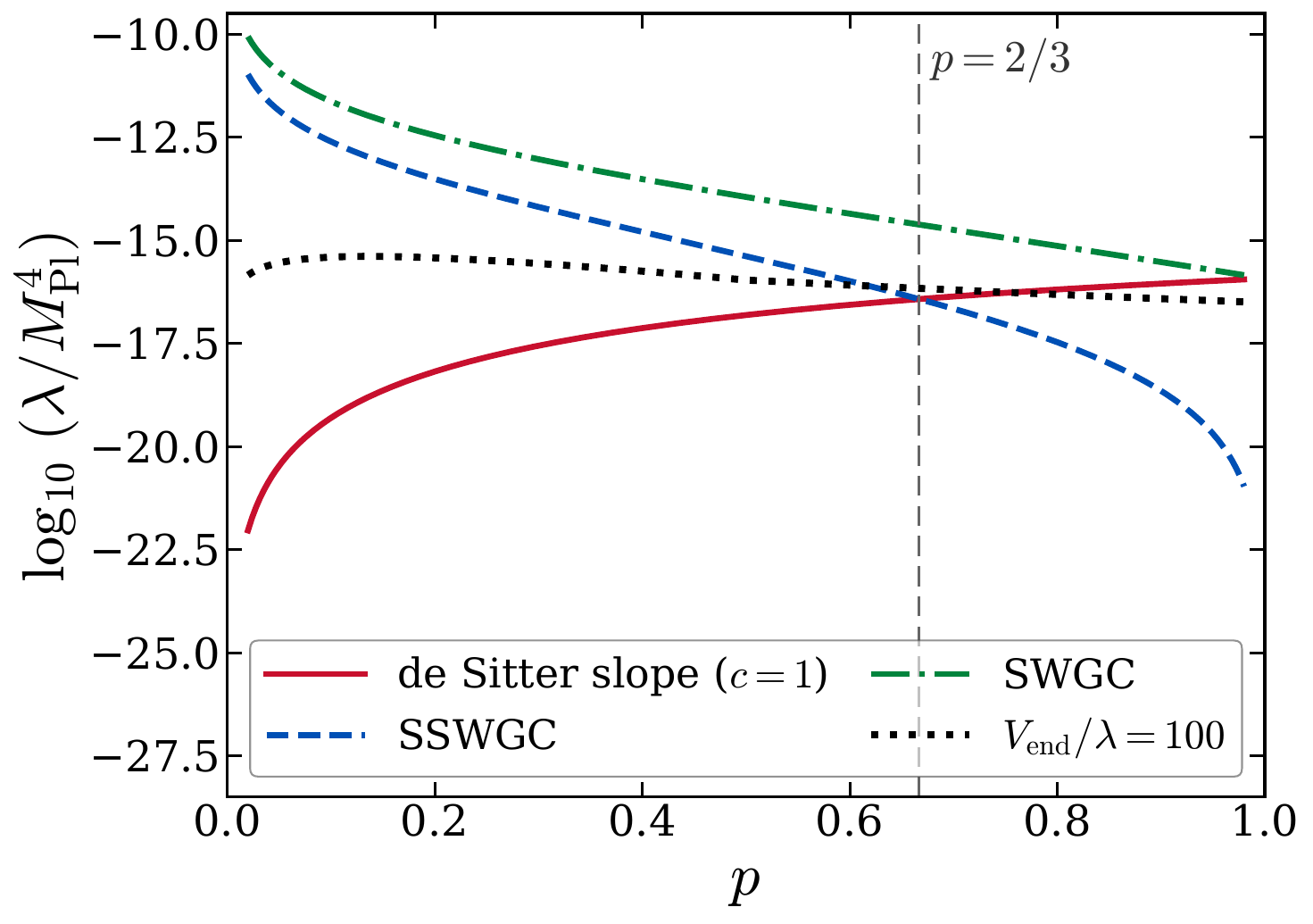}
\caption{{\it{Upper bounds on the dimensionless brane tension
$\lambda/M_{\mathrm{Pl}}^4$ for $N=60$, from the scalar-amplitude
normalization combined with the de Sitter slope criterion with $c=1$, the
SSWGC, the SWGC, and the high-energy benchmark $V_{\rm end}/\lambda=100$.
The slope criterion is the strongest limit for $p<2/3$ and the SSWGC for
$p>2/3$. The high-energy benchmark is weaker than both except close to
$p=2/3$, and the lower bound of Eq.~(\ref{eq:lambdaLowerM5}),
$\log_{10}(\lambda/M_{\mathrm{Pl}}^4)\simeq-64.3$, lies below the plotted
range.}}}
  \label{fig6}
\end{figure}

\begin{table*}[t]
\caption{Summary of the scalar Swampland diagnostics for the canonical
monomial potential. The local field bounds apply at the pivot and are
given in terms of $x_*=\phi_*/M_{\mathrm{Pl}}$. The FRDSSC depends on
the conjectural parameters $(\alpha,q)$, while the SDC applies to the
full trajectory.}
\label{tab:swampland-summary}
\centering
\begin{ruledtabular}
\begin{tabular}{llll}
Criterion
& Diagnostic
& Field-domain or parameter condition
& Status for $0<p<1$ \\
\hline
DSSC
& $F_1\geq c$
& $|x_*|\leq p/c$
& For $c=1$, the tightest local bound when $0<p<2/3$ \\[0.8ex]

FRDSSC
& $F_1^q-\alpha F_2\geq1-\alpha$
& $0<\alpha<1,\quad q>2$
& Finite interval depending on $(\alpha,q,p)$ \\[0.8ex]

SWGC
& $d_3^2-d_2^2\geq0$
& $|x_*|\leq2-p$
& Finite field interval \\[0.8ex]

SSWGC
& $2d_3^2-d_2d_4-d_2^2\geq0$
& $|x_*|\leq\sqrt{(1-p)(2-p)}$
& More restrictive than the SWGC \\[0.8ex]

SDC
& $|\Delta\phi|/M_{\mathrm{Pl}}\lesssim\mathcal{O}(1)$
& Full inflationary trajectory
& No explicit tower of light states is specified \\[0.8ex]

TCC
& $e^{N_{\rm TCC}}H_{\rm e}\lesssim M_{\mathrm{Pl}}$
& Conditional on the ultraviolet cutoff
& Holds only for $N_{\rm TCC}\lesssim33$ (Sec.~\ref{IIIC})
\end{tabular}
\end{ruledtabular}
\end{table*}

In Fig. \ref{fig6} we compare the upper bounds on the brane tension for $N=60$.
The scalar-amplitude normalization translates each local field-space
condition into an upper bound on $\lambda$, while the dotted curve enforces
the high-energy benchmark at the end of inflation. A viable value of
$\lambda$ must satisfy the most restrictive applicable upper bound and
the lower bound in Eq.~(\ref{eq:lambdaLowerM5}), which lies far below the
plotted range.

The brane tension is also subject to a lower bound through the
five-dimensional gravitational scale. In the Planck-mass convention
of Eq.~\eqref{eq:PlanckRelation}, a lower limit $M_5\geq M_{5,\min}$
implies
\begin{equation}
\lambda\geq
\frac{3M_{5,\min}^{\,6}}{4\pi M_4^2},
\label{eq:lambdaLowerM5}
\end{equation}
with $M_4$   the unreduced four-dimensional Planck mass. We mention that the
original brane analysis quotes a lower scale of roughly
$M_5\sim10^5\,\mathrm{TeV}$ from the requirement that four-dimensional
gravity should be recovered at distances above the millimeter scale
\cite{LinNgCheung2019}. With Eq.~\eqref{eq:PlanckRelation} this gives
$\lambda\gtrsim(200\,\mathrm{GeV})^4$, i.e.
$\lambda/M_{\mathrm{Pl}}^4\gtrsim5\times10^{-65}$.

Lastly, note that the resulting parameter region is determined by the overlap
between
this lower bound and the upper limits from the field-space criteria
and the high-energy condition. For $N=60$ and $p=0.15$ the most
restrictive upper limit is the de Sitter slope bound,
$\lambda/M_{\mathrm{Pl}}^4\lesssim2\times10^{-19}$, so the allowed interval
spans about 45 orders of magnitude. The TCC limit of
Eq.~\eqref{eq:lambdaTCC} falls inside this interval only for
$N_{\rm TCC}\lesssim33$.

\subsection{Joint observational and Swampland viability}

The observational and theoretical conditions can be combined by
examining the parameter dependence through the model. For each pair
$(p,N)$, the leading-order expressions in
Eqs.~\eqref{eq:nsmono},\eqref{eq:running} and (\ref{eq:rmono}) determine the
inflationary
observables. Then, the CMB constraints in Sec. \ref{CMBconstraintssec} select the
allowed
region in the $(p,N)$ plane, and the scalar-amplitude normalization fixes
the relation between the pivot field value and the brane tension
through Eq.~\eqref{eq:xstarLambda}. Finally, the Swampland conditions and
the high-energy requirement further
restrict the allowed parameters.

For the phenomenologically relevant branch $0<p<1$, the scalar
field-space conditions are summarized in
Table~\ref{tab:swampland-summary}. The local criteria apply at the
pivot field value $x_*=\phi_*/M_{\mathrm{Pl}}$, whereas the distance
conjecture constrains the full inflationary trajectory.

For $c=1$, the hierarchy
$
\frac{p}{c}
<
\sqrt{(1-p)(2-p)}
<
2-p
$
holds for $0<p<2/3$. In this range, the ordinary de Sitter slope
criterion is more restrictive than the SSWGC, while the SWGC permits
the widest field interval. For $2/3<p<1$, the SSWGC gives the
narrowest of these three local windows.

\begin{figure}[htbp]
\centering
\includegraphics[width=8.4cm]{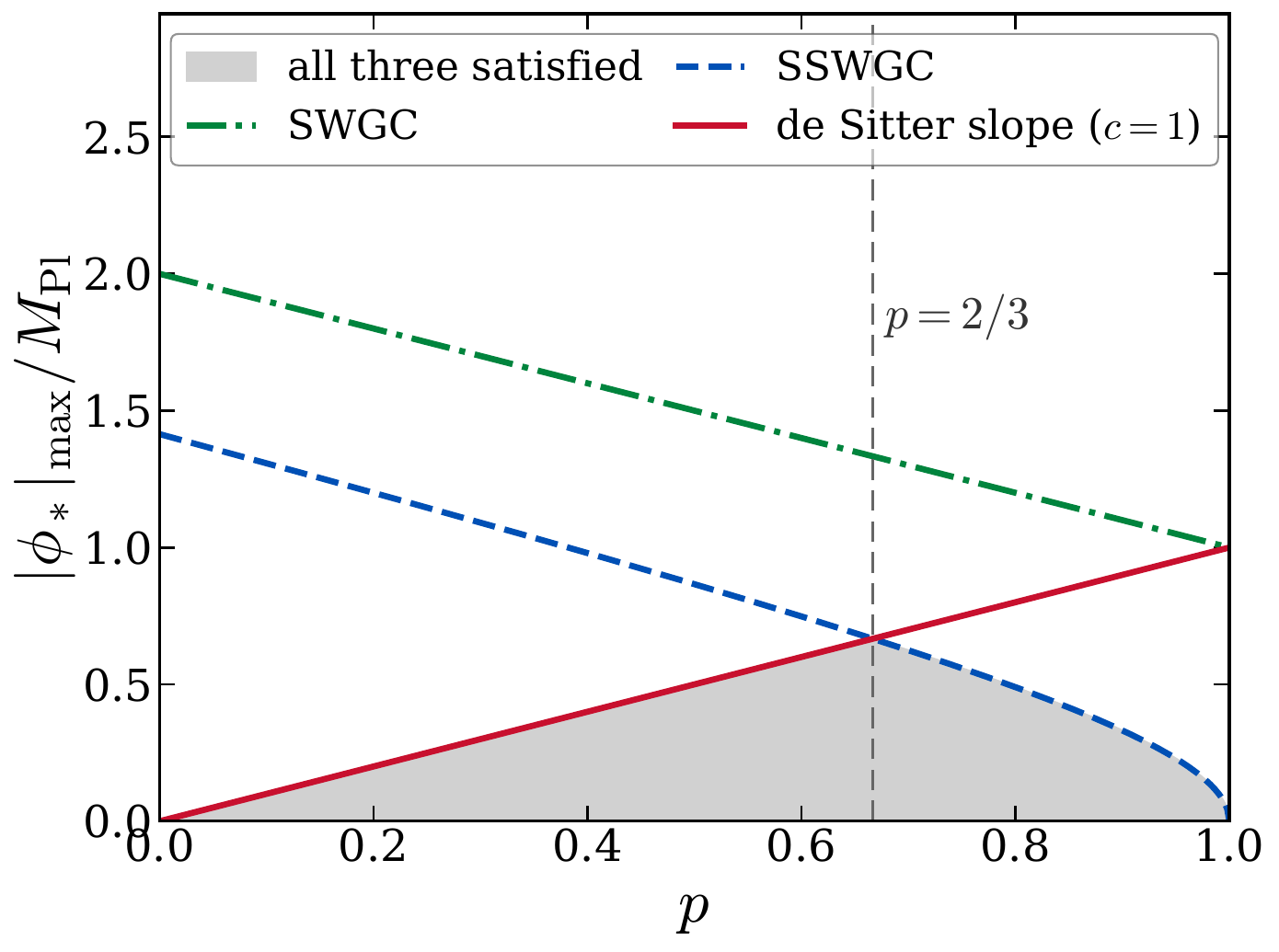}
\caption{{\it{Upper limits on the pivot field value $|\phi_*|/M_{\mathrm{Pl}}$
for the monomial potential with $0<p<1$, from the de Sitter slope criterion
with $c=1$, the SSWGC, and the SWGC. In the shaded region all three hold.
At $p=2/3$ the strongest limit passes from the de Sitter slope criterion to
the SSWGC.}}}
\label{fig3}
\end{figure}

The hierarchy of the local field-space bounds is shown in Fig.~\ref{fig3}. As we can
see, for $0<p<2/3$ the de Sitter slope criterion with $c=1$ gives the tightest
bound. For $2/3<p<1$, the SSWGC is more restrictive, while the SWGC permits
the widest field interval throughout $0<p<1$. The shaded region marks the
pivot-field values that satisfy all three conditions simultaneously.

The field windows translate into brane-tension bounds through the
normalization relation in Eq.~\eqref{eq:xstarLambda}. The corresponding
bounds are given in Eqs.~\eqref{eq:lambdaDSSC}-\eqref{eq:lambdaSSWGC},
with the distance-conjecture bound given by Eq.~\eqref{eq:lambdaSDC}.
As we mentioned, a viable point must also satisfy the high-energy condition at the
end
of inflation, namely Eq.~\eqref{eq:HEendcriterion}. The
lower bound inherited from the five-dimensional gravitational scale,
Eq.~\eqref{eq:lambdaLowerM5}, must be imposed at the same time  too, while the
TCC limit of \eqref{eq:lambdaTCC} can be added to this intersection
only in the low-$N$ corner identified in Sec.~\ref{IIIC}.

Thus, the   viability result is obtained by intersecting the CMB
allowed region in $(p,N)$ with the amplitude-normalized field windows,
the high-energy condition, and the allowed interval in $\lambda$. Note that the
CMB constraints alone do not establish simultaneous Swampland
compatibility. Likewise, satisfaction of local criteria at the pivot
does not guarantee their validity along the entire trajectory. The
resulting overlap is a finite, conditional region of the adopted
effective parameter space, and it does not establish a complete
ultraviolet-consistent quantum-gravity realization.

\section{Theoretical consistency and interpretation}
\label{Theoreticalconsistency}

The viable regions identified above rely on assumptions about
the brane background, its perturbations, and the range of validity of
the effective description. In this section, we clarify the status of these
assumptions and the interpretation of the resulting parameter space.

\subsection{Validity of the high-energy and brane-perturbation approximations}
\label{IVA}

The high-energy expressions used in this work are asymptotic
predictions. Their validity should be assessed throughout
inflation, with the end of inflation providing the most restrictive
point because the potential decreases along the rolling trajectory.
For the monomial model, Eq.~\eqref{eq:VendLambda} gives
$V_{\rm end}/\lambda$ in terms of $\phi_{\rm end}/M_{\mathrm{Pl}}$ for the
two relevant ranges of $p$. Requiring $V_{\rm end}/\lambda\gg1$ ensures that the
high-energy condition holds throughout inflation. Combined with the
scalar-amplitude normalization, this requirement gives an upper bound on the
brane tension, as discussed in Sec.~\ref{IVBnorm}.

For definiteness, we use $V/\lambda>100$ as a phenomenological
benchmark, following the original analysis
\cite{LinNgCheung2019}. Since $u=V/(2\lambda)$, this corresponds to
$u>50$. The value 100 is an operational criterion rather than a
fundamental threshold, and results close to this boundary should
be interpreted with this qualification.

The opposite limit must be checked as well, since the five-dimensional
description requires $V\ll M_5^4$. From Eqs.~\eqref{eq:PlanckRelation},
\eqref{eq:phiStar} and \eqref{eq:xstarLambda} we find
\begin{equation}
\frac{V_*}{M_5^4}
\simeq\left(\frac{3}{32\pi^2}\right)^{2/3}
\frac{2Np(p+2)}{\mathcal{K}(p,N)^{1/3}},
\label{eq:VoverM5}
\end{equation}
which does not depend on the brane tension. For $N=60$ and $p=0.15$ it
equals $5\times10^{-5}$, i.e. $V_*^{1/4}\simeq0.08\,M_5$, so the energy
density at the pivot is below the five-dimensional Planck scale, although
only by about one order of magnitude in energy.

The background equations alone do not determine the perturbation spectra. Scalar
fluctuations can source bulk gravitational modes, and their
subhorizon behavior can differ from that of standard
four-dimensional inflation \cite{KoyamaEtAl2007}. The expressions for
$n_s$ and $r$ used here are leading-order slow-roll
predictions within the effective brane description, and they do not
constitute a full five-dimensional perturbation calculation or a
likelihood analysis that includes the coupled brane-bulk system.

The Swampland inequalities are evaluated for the canonically normalized scalar field
on the brane. Without specifying the bulk geometry, moduli
stabilization, the origin of the potential normalization $A$, the
status of the brane tension, and any relevant tower of states, these
conditions do not establish that the underlying five-dimensional
model has a consistent ultraviolet completion.

The brane tension should also not be treated as an entirely free
parameter. In a more complete theory, it may be related to the bulk
curvature, the five-dimensional Planck scale, fluxes, compactification
volumes, or supersymmetry-breaking scales. Decreasing $\lambda$ changes
the amplitude-normalized field range, but may also move other
parameters outside the regime of validity of the effective theory.

\subsection{Relation to previous analyses, analytic results, and limitations}
\label{IVC}

Earlier works established that the high-energy brane correction can
suppress the tensor amplitude while allowing a steeper potential
\cite{LinNgCheung2019}. Subsequent studies extended the analysis to
reheating, inverse-monomial potentials, and the trans-Planckian
censorship conjecture
\cite{MohammadiGolanbariNasriSaaidi2022,Mohammadi:2020ctd}. More recently,
the further-refined de Sitter criterion and
scalar weak gravity conditions were applied to inverse-monomial brane inflation
\cite{NooriGashti:2024ldh}.

The present analysis is based on these
developments by relating potential derivatives to observables
through the high-energy RS-II background. These relations
depend on the brane quantity $u_*=V_*/(2\lambda)$, so $n_s$ and $r$ alone
do not determine the potential slope and curvature. A joint
compatibility assessment must also account for the scalar normalization,
the high-energy condition, the reheating history, and the ultraviolet assumptions.

The high-energy monomial solution gives two results that can be stated
analytically. At leading order, $n_s$ and $r$ depend only on $p$ and $N$ and obey
the consistency relation Eq.~(\ref{eq:nsrNidentity}).
Moreover, the SWGC and SSWGC yield
exact analytic field-space intervals. Away from the degenerate and
singular endpoints, the SWGC requires
$|\phi|/M_{\mathrm{Pl}}\leq|p-2|$, while the SSWGC gives the branches
in Eq.~\eqref{eq:SSWbranches}. These
conditions hold only over finite field intervals and do not establish
consistency across the full monomial domain.

The observational comparison identifies a broader compatible
sublinear branch for the higher-tilt ACT-era benchmark. The Planck
2018 overlap is more restricted and concentrated toward smaller
values of $N$. In both cases, the allowed region depends on the post-inflationary
history and the validity of the high-energy approximation.

Finally, we mention two additional limitations. The model does
not specify the tower of states needed to assess the full distance
conjecture. The TCC is the most restrictive criterion, i.e. with
$N_{\rm TCC}\geq N_*$ and the lower bound of Eq.~\eqref{eq:lambdaLowerM5} it
holds only for $N_*\lesssim33$ and $p\lesssim0.04$, outside the benchmark
range $N=50$--$70$, and thus a five-dimensional ultraviolet scale would
strengthen the requirement.

\section{Conclusions}
\label{Conclusions}

High-energy inflation on an RS-II brane offers a natural framework in which one can
examine
the relation between early-Universe observations and conjectured quantum-gravity
constraints. The quadratic energy-density correction changes the slow-roll dynamics
and may allow a monomial potential to satisfy observational bounds while occupying a
restricted field-space region. This possibility is of particular interest given the
different regions preferred by Planck and the higher-tilt ACT-era data. We have
reassessed this question for the potential $V(\phi)=A\phi^p$, keeping the assumptions
of the high-energy brane description explicit throughout.

A central result is that the usual four-dimensional relations between the potential
derivatives and the CMB observables do not hold in the high-energy RS-II regime. The
corresponding brane expressions depend on $u_*=V_*/(2\lambda)$, thus neither the
tensor
amplitude alone nor the scalar tilt and tensor amplitude together determine universal
values of the potential slope and curvature. The monomial model nevertheless yields
the simple leading-order relation $n_s\simeq1-1/N-r/8$. This relation provides a
direct observational test of the high-energy scenario, while making clear that the
CMB comparison remains conditional on the e-fold number and the post-inflationary
history.

The scalar Swampland diagnostics lead to finite analytic field-space windows. For
$0<p<1$, the SWGC permits $|\phi|/M_{\rm Pl}\leq 2-p$, while the SSWGC gives the
narrower interval $|\phi|/M_{\rm Pl}\leq\sqrt{(1-p)(2-p)}$. For the ordinary de
Sitter criterion with $c=1$, the slope bound is more restrictive for $0<p<2/3$,
whereas the SSWGC becomes tighter for $2/3<p<1$. The negative curvature of the
potential in this sublinear branch can also assist the further-refined de Sitter
condition. These results identify portions of field space where the three local
diagnostics
can hold together, i.e. they do not imply global satisfaction along the full field
domain
or establish a quantum-gravity completion.

The CMB comparison still allows for a region of parameter space. In
particular, the higher-tilt ACT-era benchmark permits a broader sublinear branch,
while the
Planck-compatible region is more restricted and concentrated toward lower values of
$N$ in the leading-order treatment. The scalar-amplitude normalization translates the
field-space conditions into bounds on the brane tension, which must be considered
together with the high-energy requirement at the end of inflation and any independent
lower bound from the five-dimensional gravitational scale. A sub-Planckian field
excursion can be obtained, however the full distance conjecture cannot be assessed
without specifying the associated tower of states. The TCC is far more restrictive
than the local
criteria, namely  once the accelerated phase lasts at least as long as the
observable e-folds
and the brane tension respects the limit from the five-dimensional gravitational
scale, it holds only for about 32 e-folds and $p\lesssim0.04$, although with a
tensor amplitude that need not be negligible. Thus, the
resulting
overlap should be understood as phenomenological compatibility with the local
criteria within the adopted effective description, rather than evidence for a
complete ultraviolet
realization.

We close this work by mentioning  that a
full five-dimensional
treatment of scalar and tensor perturbations, including bulk Weyl effects and
Kaluza-Klein modes, could strengthen the comparison with current and future CMB
data. Furthermore, a microscopic construction specifying the origin of the potential,
the stabilization
of the brane tension, and the relevant spectrum of states, would allow more direct
tests of the Swampland criteria. Moreover, more detailed reheating analyses for
fractional powers
$0<p<1$ and joint likelihood studies using the Planck, ACT, BICEP/Keck, and DESI
data,
could sharpen the allowed range of $N$ and the model parameters. Lastly,
extensions to other
potentials and multifield brane models, alongside a formulation of the TCC in a
genuinely five-dimensional setting, could help to distinguish features of the
monomial
model from those of braneworld inflation more generally. These studies lie beyond
the scope of the present work and are left for future projects.

 \begin{acknowledgments}

ENS acknowledges the contributions of the LISA
CosWG  and  COST
Actions
 CA21106 ``COSMIC WISPers
in the Dark Universe: Theory, astrophysics and experiments'',  CA21136
``Addressing observational tensions in cosmology with
  systematics and fundamental physics (CosmoVerse)'',    CA23130
``Bridging high and low energies in
search of quantum gravity (BridgeQG)'', and CA24101 ``Testing Fundamental
Physics with Seismology''. \.{I}.S. is grateful to EMU, T\"{U}B\.{I}TAK, ANKOS, and
SCOAP3 for their support, and acknowledges the networking support of COST Actions
CA18108, CA22113, CA21106, CA23130, CA23115, and CA21136.

 \end{acknowledgments}

\end{document}